\documentclass{article}
\usepackage{spconf,amsmath,graphicx,booktabs}
\usepackage{newtxtext,newtxmath}
\usepackage[hidelinks]{hyperref}

\usepackage{etoolbox}
\newcommand{\tightdisplay}{\setlength{\abovedisplayskip}{5pt plus 2pt minus 2pt}\setlength{\belowdisplayskip}{5pt plus 2pt minus 2pt}\setlength{\abovedisplayshortskip}{2pt plus 2pt}\setlength{\belowdisplayshortskip}{3pt plus 2pt minus 1pt}}
\appto\normalsize{\tightdisplay}
\title{An offset--drift Kalman filter for continuous emotion recognition\\and annotation from sparse human reviews}

\name{Jie Dong, Mahdi Zareei\sthanks{Corresponding author.}}
\address{School of Engineering and Sciences, Tecnologico de Monterrey, Monterrey, NL, 64700, Mexico}

\begin{document}
\maketitle

\begin{abstract}
Continuous emotion recognition tracks how emotion changes over time, but its predictions often deviate from human ratings, and the traces it is trained on are rated in real time by several annotators, which is slow and costly. A few human reviews could help with both, but rejection only decides which predictions to hand over and adaptation adjusts the model or its outputs; neither uses reviews to correct unreviewed windows directly. We propose a Kalman-filter correction that models the predictor's error as a per-part offset plus a slow drift, estimates it from the reviewed windows and corrects every unreviewed window online, leaving the predictor unchanged. On MSP-Conversation, it beats four simpler corrections in all 48 paired comparisons, keeps correcting each part after reviewing stops and, for uncorrected predictions, tells a person which windows to take over. The same correction serves annotation: with 20\% of windows reviewed, the completed trace comes closer to the raters.
\end{abstract}

\begin{keywords}
continuous emotion recognition, continuous emotion annotation, bias correction, Kalman filter, human review
\end{keywords}

\section{Introduction}
\label{sec:intro}

\begin{figure}[t]
\centering
\includegraphics[width=\columnwidth]{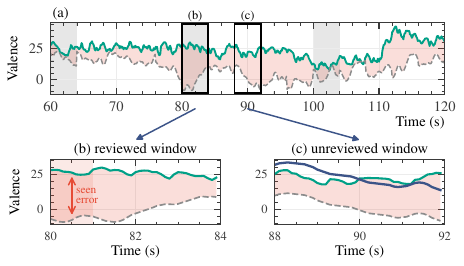}
\caption{(a) A 60\,s excerpt of valence (original rating scale) in one development part: the raters' mean (green), the uncorrected prediction (grey dashed) and the signed error between them (shaded); grey bands: reviewed windows (equal interval, 20\%). (b) A reviewed window; red: its first second, the only part available in time. (c) An unreviewed window after it, with the ODKF correction (blue). The part has the median absolute mean error.}
\label{fig:trace}
\end{figure}

Continuous emotion recognition describes emotion as values that change over time, since emotion itself evolves continuously. A recognition system both outputs such traces and learns from them: its training data are traces for which several annotators listen to a recording and move a slider while it plays, with their ratings averaged into one trace per dimension~\cite{ringeval2013recola,martinezlucas2020conversation}. However, neither kind of trace is easy to obtain. The predicted trace carries errors that are not independent from moment to moment: the prediction stays on one side of the raters' mean for long stretches, as in Fig.~\ref{fig:trace}(a). The annotated trace is costly, because ratings differ enough between people that several raters are needed per recording~\cite{wu2025raters}. Both can be improved in the same way: people review a small part of the trace, and their reviews are used to improve the whole trace.

Two lines of work bear on this setting. The first improves the predictor itself. Models of emotion dynamics describe the rating trace as evolving continuously, with Kalman filters~\cite{dang2017kalman,huang2017epps}, sequential Monte Carlo~\cite{wu2022smc} or constrained neural ordinary differential equations~\cite{dang2023cdnode,wu2024dcnode}; they model the emotion itself, not the predictor's error. Adaptation methods retrain the model, or adapt it or its outputs at test time, for new speakers or test data~\cite{abdelwahab2018domain,sridhar2022unsupervised,dong2026tta}, and Tran et al.\ recalibrate each test speaker's predictions without target labels, mapping their mean and spread onto the label statistics of the most similar training speakers~\cite{tran2023personalized}. None of these uses human input at test time. The second line brings people in. Rejection and uncertainty estimation decide which predictions a person should check~\cite{sridhar2019reject,sridhar2020uncertainty,li2022confidence,wu2023deer}; active and cooperative learning, including active test-time adaptation~\cite{gui2024active}, use people's labels to train or adapt the model~\cite{zhang2015cooperative,abdelwahab2017incremental}; and in human-assisted transcription each correction stays in its segment~\cite{sperber2013respeaking,sperber2014segmentation}. An answer reaches further only when the output has a structure to carry it, as when an answer in human-assisted speaker diarization merges or splits a whole branch of a clustering tree~\cite{shamsi2022diarization}. A continuous trace has such a structure in time: a predictor that misjudges a stretch of speech carries the same bias over many windows, so a reviewed window carries information about its unreviewed neighbours. However, we found no work that uses this structure to correct the predictor's output online.

This paper frames human review of a continuous emotion trace as a tracking problem, in which reviewed windows are observations of the predictor's error. We first analyse, on MSP-Conversation~\cite{martinezlucas2020conversation}, the errors of a neural-ODE predictor relative to the raters' mean; this analysis reveals a constant offset per recording part plus a slow drift. Building on this structure, we propose an offset--drift Kalman filter (ODKF), a state-space correction whose observation model follows how people rate a trace and which leaves the predictor unchanged. In a closed-loop evaluation, the correction beats four simpler corrections, including an estimate of the part offset alone, in all 48 paired comparisons. Its estimate also outlives the reviews: when reviewing stops a quarter, half or three quarters of the way through each part, it keeps lowering the loss on the remaining windows, and if the predictions are left uncorrected, it tells a person which windows to take over. Finally, the same correction benefits annotation: the completed trace comes closer to the raters, most of all when the reviews are spread evenly along it.

\section{Offset--drift Kalman filter}
\label{sec:method}

We frame the correction as Bayesian tracking: in each recording part, the predictor's error on every window is inferred from earlier reviews and added to the prediction. It uses only the predictor's output and the reviews, so it does not depend on the input modality; we evaluate it on speech. The model of that error, and of what a review reveals about it, follows how people rate a trace: a part is rated as a whole by one group of raters, the raters respond with a lag, and a predictor's bias depends on the part's speakers and raters.

\subsection{Problem setting}
\label{ssec:problem}

A recording part is rated by $M$ annotators at 10\,Hz; $\bar y_t$ is the mean of their ratings at frame $t$ and $s_t^2$ their sample variance. A predictor gives $f_t$ for every frame before any label is seen. The part is cut into non-overlapping 4\,s windows (40 frames), $k = 1, \dots, n$. A reviewed window takes the raters' mean in the final trace. The error of window $k$ is
\begin{equation}
e_k = \frac{1}{40} \sum_{t \in k} \left( \bar y_t - f_t \right).
\label{eq:err}
\end{equation}
Because the released traces are shifted back by 3\,s to compensate for the raters' reaction lag, only the first second of a reviewed window $j$ has been rated before the next window must be predicted. The observation is therefore
\begin{equation}
o_j = \frac{1}{10} \sum_{t \in U_j} \left( \bar y_t - f_t \right),
\label{eq:obs}
\end{equation}
where $U_j$ holds the first 10 frames of window $j$.

\subsection{Error model}
\label{ssec:model}

We write the error of window $k$ in part $p$ as a part offset plus a drift, and the observation as the error plus noise:
\begin{align}
e_k &= a_p + b_k, \label{eq:sum}\\
a_p &\sim \mathcal{N}(0, \tau^2), \quad b_1 \sim \mathcal{N}(0, q), \label{eq:offset}\\
b_k &= \phi\, b_{k-1} + \eta_k, \quad \eta_k \sim \mathcal{N}\!\left(0, (1-\phi^2) q\right), \label{eq:ar}\\
o_j &= e_j + \xi_j, \quad \xi_j \sim \mathcal{N}(0, \rho). \label{eq:obsnoise}
\end{align}
The offset $a_p$ captures that a predictor tends to over- or under-estimate a whole part; since the reference is the mean of the part's raters, it also includes that group's own constant deviation. The drift $b_k$ is a stationary first-order autoregressive process with variance $q$, for errors that follow the changing content of the conversation, persist for some seconds and then fade. Its time scale and the offset's share of the error variance are
\begin{equation}
T = -\frac{\Delta}{\ln \phi}, \qquad \pi_a = \frac{\tau^2}{\tau^2 + q},
\label{eq:timescale}
\end{equation}
with $\Delta = 4$\,s. The noise $\rho$ covers the difference between a window's first second and the whole window. Because each part has its own speakers and raters, each part starts afresh from the priors~\eqref{eq:offset}.

\subsection{Online correction}
\label{ssec:online}

With state $\mathbf{x}_k = (a_p, b_k)^\top$, the model is a linear Gaussian state-space model~\cite{kalman1960,durbin2012},
\begin{equation}
\mathbf{F} = \begin{pmatrix} 1 & 0 \\ 0 & \phi \end{pmatrix}, \quad
\mathbf{Q} = \begin{pmatrix} 0 & 0 \\ 0 & (1-\phi^2) q \end{pmatrix}, \quad
\mathbf{h} = \begin{pmatrix} 1 \\ 1 \end{pmatrix},
\label{eq:ssm}
\end{equation}
started at $\hat{\mathbf{x}}_{1|0} = \mathbf{0}$ and $\mathbf{P}_{1|0} = \mathrm{diag}(\tau^2, q)$ in every part. The standard Kalman filter runs over the windows; it updates only at reviewed windows and otherwise carries its prediction forward. The corrected prediction of window $k$ is
\begin{equation}
\tilde y_t = f_t + \mathbf{h}^\top \hat{\mathbf{x}}_{k|k-1}, \quad t \in k,
\label{eq:corr}
\end{equation}
which uses only reviews before $k$. Online correction suits review in playback order; offline, once all reviews are in, the Rauch--Tung--Striebel smoother~\cite{rauch1965} also uses later reviews.

\begin{table*}[t]
\centering
\caption{Loss~\eqref{eq:fpc} on unreviewed test windows ($\times 10^{-4}$, three seeds): the uncorrected level, then each method's difference to it. Bold: significantly better than every baseline.}
\label{tab:main}
\footnotesize
\setlength{\tabcolsep}{3.2pt}
\begin{tabular}{@{}lcccccccccccc@{}}
\toprule
 & \multicolumn{6}{c}{Arousal} & \multicolumn{6}{c}{Valence} \\
\cmidrule(lr){2-7}\cmidrule(l){8-13}
 & \multicolumn{3}{c}{Random} & \multicolumn{3}{c}{Equal interval} & \multicolumn{3}{c}{Random} & \multicolumn{3}{c}{Equal interval} \\
\cmidrule(lr){2-4}\cmidrule(lr){5-7}\cmidrule(lr){8-10}\cmidrule(l){11-13}
Method & 5\% & 10\% & 20\% & 5\% & 10\% & 20\% & 5\% & 10\% & 20\% & 5\% & 10\% & 20\% \\
\midrule
Uncorrected & 4.71 & 4.68 & 4.71 & 4.68 & 4.70 & 4.72 & 46.07 & 46.01 & 45.80 & 46.21 & 46.16 & 46.41 \\
\midrule
LAST & $+1.45$ & $-0.01$ & $-2.40$ & $+5.49$ & $+0.54$ & $-3.57$ & $-5.40$ & $-11.90$ & $-18.94$ & $-1.17$ & $-14.50$ & $-24.46$ \\
CM & $-1.85$ & $-4.88$ & $-7.54$ & $+3.24$ & $-3.69$ & $-7.75$ & $-10.06$ & $-17.17$ & $-22.22$ & $-7.23$ & $-18.87$ & $-24.68$ \\
EWMA & $-1.64$ & $-5.05$ & $-8.31$ & $+2.55$ & $-4.76$ & $-9.12$ & $-9.34$ & $-18.07$ & $-26.32$ & $-3.89$ & $-21.04$ & $-30.94$ \\
Offset-only & $-5.37$ & $-7.14$ & $-8.86$ & $-4.52$ & $-7.28$ & $-9.33$ & $-15.03$ & $-20.20$ & $-23.72$ & $-14.26$ & $-21.55$ & $-25.55$ \\
\midrule
ODKF (ours) & $\mathbf{-5.60}$ & $\mathbf{-7.62}$ & $\mathbf{-9.72}$ & $\mathbf{-4.94}$ & $\mathbf{-7.94}$ & $\mathbf{-10.60}$ & $\mathbf{-16.50}$ & $\mathbf{-23.25}$ & $\mathbf{-29.15}$ & $\mathbf{-16.38}$ & $\mathbf{-25.68}$ & $\mathbf{-32.64}$ \\
\bottomrule
\end{tabular}
\end{table*}

\subsection{Parameter estimation}
\label{ssec:params}

The parameters $\tau^2$, $\phi$ and $q$ are fitted on the development set, where every window can be treated as observed. Under~\eqref{eq:sum}--\eqref{eq:ar}, the errors of a part with $n$ windows are jointly Gaussian,
\begin{equation}
\mathbf{e}_p \sim \mathcal{N}\!\left( \mathbf{0},\; \tau^2 \mathbf{1}\mathbf{1}^\top + q\, \boldsymbol{\Phi} \right), \quad \Phi_{ij} = \phi^{|i-j|},
\label{eq:lik}
\end{equation}
and we maximise the summed log-likelihood over development parts with L-BFGS from three fixed starting points, with a free mean that the correction sets to zero; $\rho$ is the development mean of $(o_j - e_j)^2$. Development experiments use parameters cross-fitted over five folds of parts; the test set uses one fit on all development parts. Across the three predictor seeds, the offset carries 0.48--0.49 of the error variance for arousal and 0.43--0.47 for valence, and the drift has a time scale of 10.5--10.9\,s and 20.6--21.9\,s.

\section{Experimental settings}
\label{sec:setup}

\subsection{Data}
\label{ssec:data}

MSP-Conversation~\cite{martinezlucas2020conversation} (version 2.0) contains podcast conversations cut into parts and rated continuously for arousal and valence at 10\,Hz, by six to eight annotators per part. Its labelled portion has 799 parts from 279 conversations (68.3\,h). The official speaker-independent partition, whose test speakers are unseen in training, has 484 parts for training, 147 for development and 168 for the first test set. The released traces are lag-compensated~\cite{mariooryad2015delay} and median-filtered; we use them as released. Ratings are mapped linearly from $[-100, 100]$ to $[0, 1]$, and all losses are on this scale.

\subsection{System configuration}
\label{ssec:predictor}
\label{ssec:protocol}

\textbf{Predictor.} Our predictor is a neural ODE~\cite{chen2018neuralode} with the rate constraint of CD-NODE~\cite{dang2023cdnode} and, from its dual-constrained extension~\cite{wu2024dcnode}, the range constraint and a second ODE-driven line that models the rater spread alongside the mean. Its input is the average of the 25 hidden layers of a frozen WavLM-Large~\cite{chen2022wavlm}, pooled over each label frame. A fixed-step fourth-order Runge--Kutta solver integrates at 0.1\,s. It is trained on the training set for each dimension with three seeds (42, 123, 456). A second network of the same form, trained on the development set (cross-fitted over five folds there, fold models averaged on the test set), adds a window-level refinement of the mean. The result is the uncorrected prediction $f_t$, shared by all methods.

\textbf{Review protocol.} A review reveals the raters' mean on a window, which is also the reference for every window. Following Wu et al., whose results on this corpus suggest that five raters may suffice for a stable estimate of the rating distribution~\cite{wu2025raters}, a review should come from at least five raters; every part here has six to eight, and each review uses all of them. Two schedules, neither of which looks at the predictions, choose the reviewed windows: random review, averaged over ten draws, and equal-interval review of every $m$-th window from the first. Budgets are 5, 10 and 20\% of windows; the labelling effort is taken as proportional to the number of reviewed windows. In a further setting, equal-interval review stops after the first quarter, half or three quarters of each part; at half, and with no review at all, 20\% of the remaining windows are handed over to a human. A sensitivity check raises the reaction gap from 3\,s to 3.5\,s, which leaves 5 usable frames per review.

\subsection{Baselines}
\label{ssec:baselines}

All baselines add a correction $\hat e_k$ to $f_t$ as in~\eqref{eq:corr}, computed from the same observations of earlier reviewed windows. LAST carries the most recent observation forward; CM averages all observations so far; EWMA weights observation $j$ by
\begin{equation}
w_j = 2^{-(t_{\mathrm{last}} - t_j)/h},
\label{eq:ewma}
\end{equation}
where $t_j$ is the time of review $j$ and $t_{\mathrm{last}}$ that of the latest, with the half-life $h$ chosen on the development set from $\{3, 6, \dots, 96\}$\,s (48\,s for arousal, 24\,s for valence); and Offset-only, a shrinkage estimate of the part offset alone, is the model of Section~\ref{ssec:model} with $\phi = 0$, fitted in the same way.

\section{Results and analysis}
\label{sec:results}

\subsection{Evaluation methodology}
\label{ssec:eval}

\textbf{Loss.} The main measure is the mean loss over unreviewed windows. The squared error against the mean of $M_t$ raters contains that mean's sampling variance, which in this corpus is of the same order as the variation of the trace itself. Treating the consensus as an estimate from a finite group of raters~\cite{burmania2017raters,wu2025raters}, we remove this variance with a finite population correction~\cite{cochran1977} for the pool of $N = 25$ raters,
\begin{equation}
\ell_t = (\hat y_t - \bar y_t)^2 - \left(1 - \frac{M_t}{N}\right) \frac{s_t^2}{M_t},
\label{eq:fpc}
\end{equation}
where $\hat y_t$ is the prediction under evaluation. Because the reviews and the reference share raters, a corrected loss can fall below zero. The correction term depends only on the labels, so differences between methods on the same windows are unaffected; we report differences, in units of $10^{-4}$. The completed trace, with the raters' mean on reviewed windows, is also scored by the concordance correlation coefficient (CCC)~\cite{lin1989ccc}, per part (averaged) and pooled over parts.

\textbf{Significance.} Differences are paired, and a difference is called significant when, for each of the three predictor seeds, the 95\% interval of a cluster bootstrap over groups of conversations linked by a shared speaker~\cite{field2007bootstrap} (26 clusters on the test set) excludes zero.

\subsection{Experimental results}
\label{ssec:baseline-res}

The correction lowers the loss on unreviewed windows in all 12 conditions and is significantly better than each of the four baselines in all 48 comparisons (Table~\ref{tab:main}). LAST, CM and EWMA treat each review as an equally good guide and, with sparse reviews, chase its noise (for arousal at 5\% with equal intervals, all three are worse than no correction); Offset-only, which assumes one constant offset per part, is the strongest baseline except for valence at 20\%, where EWMA follows the drift. Only ODKF handles both regimes, because it weighs each review by how much of its error is a lasting offset and how much is drift that has since decayed.

Where the gain comes from follows from the same two components. Offset-only reaches 0.78 to 0.91 of the ODKF gain at 20\%: most of what a review reveals about its neighbours is where the predictor sits for the whole part, which, unlike the drift, never fades. The drift term adds the rest. ODKF is significantly better than Offset-only in all 12 conditions, by a margin that grows with the budget, as denser reviews leave less time for the drift estimate to decay.

Even with no review at all, the predictor's own spread estimate tells which valence windows to hand over, leaving a lower loss on the rest than random hand-off. With reviews, the gain outlasts the reviewing itself. Whether reviewing stops after the first quarter, half or three quarters of each part (a median of four reviews per part at a quarter), adding the filter's last estimate to the remaining windows still lowers their loss in both dimensions, as the offset estimate persists to the end of the part. If, after reviewing stops halfway, the remaining windows are left uncorrected, the filter's estimate also tells which of them to hand over: ranking them by the filter's predicted squared error, which becomes nearly constant within a part once the drift estimate has faded, hands over in effect the rest of the parts with the largest estimated offset, and leaves a lower loss than random hand-off in both dimensions.

For annotation, the completed trace matches the raters more closely (Table~\ref{tab:ccc}). With 20\% of windows reviewed at equal intervals, the per-part CCC rises from 0.47 to 0.58 for arousal and from 0.33 to 0.52 for valence.

\begin{table}[t]
\centering
\caption{CCC of the completed test trace at 20\% of windows reviewed (three seeds), per part (averaged) and pooled. Reviewed windows carry the raters' mean in every row. R: random; E: equal interval. Bold: highest in each column.}
\label{tab:ccc}
\footnotesize
\setlength{\tabcolsep}{2.5pt}
\begin{tabular}{@{}lcccccccc@{}}
\toprule
 & \multicolumn{4}{c}{Arousal} & \multicolumn{4}{c}{Valence} \\
\cmidrule(lr){2-5}\cmidrule(l){6-9}
 & \multicolumn{2}{c}{Per part} & \multicolumn{2}{c}{Pooled} & \multicolumn{2}{c}{Per part} & \multicolumn{2}{c}{Pooled} \\
\cmidrule(lr){2-3}\cmidrule(lr){4-5}\cmidrule(lr){6-7}\cmidrule(l){8-9}
Method & R & E & R & E & R & E & R & E \\
\midrule
Uncorrected & 0.463 & 0.470 & 0.624 & 0.627 & 0.326 & 0.327 & 0.640 & 0.638 \\
\midrule
LAST & 0.505 & 0.539 & 0.752 & 0.773 & 0.457 & 0.504 & 0.816 & 0.841 \\
CM & 0.519 & 0.533 & 0.777 & 0.785 & 0.407 & 0.421 & 0.809 & 0.819 \\
EWMA & 0.539 & 0.560 & 0.789 & 0.803 & 0.464 & 0.500 & 0.836 & 0.856 \\
Offset-only & 0.530 & 0.544 & 0.776 & 0.786 & 0.408 & 0.420 & 0.806 & 0.813 \\
\midrule
ODKF (ours) & \textbf{0.556} & \textbf{0.584} & \textbf{0.790} & \textbf{0.808} & \textbf{0.484} & \textbf{0.522} & \textbf{0.837} & \textbf{0.857} \\
\bottomrule
\end{tabular}
\end{table}

The error model also tells where to place the reviews. Once the predictions are corrected, reviewing at equal intervals leaves a lower loss than reviewing at random, significantly so at 20\% for both dimensions ($-0.87$ for arousal, $-2.87$ for valence) and at 10\% for valence ($-2.29$), with no condition favouring random review. The advantage comes from the correction, not from which windows remain unreviewed: without correction the two schedules leave nearly the same loss, and the difference in differences is significant in the same three conditions.

These findings hold under stricter conditions. A longer reaction gap of 3.5\,s leaves the gain in place, and offline, the smoother lowers the loss further in all 12 conditions, so the online results are the conservative case. A harder question is whom the reviews come from. With six to eight raters per part, the reviews and the reference come from the same group, so part of the gain may come from learning that group's own deviation; a single reviewer or disjoint halves of the group would be too noisy to separate the two. Resampling reviewer and answer groups of the part's size independently, with replacement, from each part's raters makes every review noisier, and the gain shrinks but stays significant in both dimensions on the development set.

\section{Conclusion}
\label{sec:conclusion}

We proposed ODKF, a state-space correction that models a continuous emotion predictor's error as a part offset plus a slow drift and turns a few reviewed windows into a correction for the rest of the trace. For recognition, the reviews lowered the loss on unreviewed windows in every condition and kept correcting each part after reviewing stopped; where the predictions were left uncorrected, the filter's estimate told a person which windows to take over. For annotation, a trace in which people review 20\% of the windows and ODKF completes the rest came closer to the raters, with the per-part CCC rising from 0.47 to 0.58 for arousal and from 0.33 to 0.52 for valence, and evenly spaced reviews completed it better than random ones.

\section{Acknowledgements}
Data provided by The University of Texas at Dallas through the Multimodal Signal Processing Lab. This material is based upon work supported by the National Science Foundation under Grants No. CNS-1823166 and CNS: 2016719. Any opinions, findings, and conclusions or recommendations expressed in this material are those of the author(s) and do not necessarily reflect the views of the National Science Foundation or The University of Texas at Dallas.

\bibliographystyle{IEEEbib}
\bibliography{refs}

\begin{thebibliography}{10}

\bibitem{ringeval2013recola}
Fabien Ringeval, Andreas Sonderegger, Juergen Sauer, and Denis Lalanne,
\newblock ``Introducing the {RECOLA} multimodal corpus of remote collaborative
  and affective interactions,''
\newblock in {\em Proc. 10th IEEE International Conference and Workshops on
  Automatic Face and Gesture Recognition (FG)}, 2013, pp. 1--8.

\bibitem{martinezlucas2020conversation}
Luz Martinez-Lucas, Mohammed Abdelwahab, and Carlos Busso,
\newblock ``The {MSP-Conversation} corpus,''
\newblock in {\em Proc. Interspeech 2020}, 2020, pp. 1823--1827.

\bibitem{wu2025raters}
Jingyao Wu, Ting Dang, Vidhyasaharan Sethu, and Eliathamby Ambikairajah,
\newblock ``How many raters do we need? {A}nalyses of uncertainty in estimating
  ambiguity-aware emotion labels,''
\newblock {\em IEEE Transactions on Affective Computing}, 2025,
\newblock Early access.

\bibitem{dang2017kalman}
Ting Dang, Vidhyasaharan Sethu, and Eliathamby Ambikairajah,
\newblock ``Dynamic multi-rater {G}aussian mixture regression incorporating
  temporal dependencies of emotion uncertainty using {K}alman filters,''
\newblock in {\em Proc. IEEE International Conference on Acoustics, Speech and
  Signal Processing (ICASSP)}, 2018, pp. 4929--4933.

\bibitem{huang2017epps}
Zhaocheng Huang and Julien Epps,
\newblock ``An investigation of emotion dynamics and {K}alman filtering for
  speech-based emotion prediction,''
\newblock in {\em Proc. Interspeech 2017}, 2017, pp. 3301--3305.

\bibitem{wu2022smc}
Jingyao Wu, Ting Dang, Vidhyasaharan Sethu, and Eliathamby Ambikairajah,
\newblock ``A novel sequential {M}onte {C}arlo framework for predicting
  ambiguous emotion states,''
\newblock in {\em Proc. IEEE International Conference on Acoustics, Speech and
  Signal Processing (ICASSP)}, 2022, pp. 8567--8571.

\bibitem{dang2023cdnode}
Ting Dang, Antoni Dimitriadis, Jingyao Wu, Vidhyasaharan Sethu, and Eliathamby
  Ambikairajah,
\newblock ``Constrained dynamical neural {ODE} for time series modelling: A
  case study on continuous emotion prediction,''
\newblock in {\em Proc. IEEE International Conference on Acoustics, Speech and
  Signal Processing (ICASSP)}, 2023.

\bibitem{wu2024dcnode}
Jingyao Wu, Ting Dang, Vidhyasaharan Sethu, and Eliathamby Ambikairajah,
\newblock ``Dual-constrained dynamical neural {ODE}s for ambiguity-aware
  continuous emotion prediction,''
\newblock in {\em Proc. Interspeech 2024}, 2024, pp. 3185--3189.

\bibitem{abdelwahab2018domain}
Mohammed Abdelwahab and Carlos Busso,
\newblock ``Domain adversarial for acoustic emotion recognition,''
\newblock {\em IEEE/ACM Transactions on Audio, Speech, and Language
  Processing}, vol. 26, no. 12, pp. 2423--2435, 2018.

\bibitem{sridhar2022unsupervised}
Kusha Sridhar and Carlos Busso,
\newblock ``Unsupervised personalization of an emotion recognition system: The
  unique properties of the externalization of valence in speech,''
\newblock {\em IEEE Transactions on Affective Computing}, vol. 13, no. 4, pp.
  1959--1972, 2022.

\bibitem{dong2026tta}
Jiaheng Dong, Hong Jia, and Ting Dang,
\newblock ``Test time adaptation for speech emotion recognition,''
\newblock in {\em Proc. IEEE International Conference on Acoustics, Speech and
  Signal Processing (ICASSP)}, 2026, pp. 17157--17161.

\bibitem{tran2023personalized}
Minh Tran, Yufeng Yin, and Mohammad Soleymani,
\newblock ``Personalized adaptation with pre-trained speech encoders for
  continuous emotion recognition,''
\newblock in {\em Proc. Interspeech}, 2023, pp. 636--640.

\bibitem{sridhar2019reject}
Kusha Sridhar and Carlos Busso,
\newblock ``Speech emotion recognition with a reject option,''
\newblock in {\em Proc. Interspeech 2019}, 2019, pp. 3272--3276.

\bibitem{sridhar2020uncertainty}
Kusha Sridhar and Carlos Busso,
\newblock ``Modeling uncertainty in predicting emotional attributes from
  spontaneous speech,''
\newblock in {\em Proc. IEEE International Conference on Acoustics, Speech and
  Signal Processing (ICASSP)}, 2020, pp. 8384--8388.

\bibitem{li2022confidence}
Yang Li, Constantinos Papayiannis, Viktor Rozgic, Elizabeth Shriberg, and Chao
  Wang,
\newblock ``Confidence estimation for speech emotion recognition based on the
  relationship between emotion categories and primitives,''
\newblock in {\em Proc. IEEE International Conference on Acoustics, Speech and
  Signal Processing (ICASSP)}, 2022, pp. 7352--7356.

\bibitem{wu2023deer}
Wen Wu, Chao Zhang, and Philip Woodland,
\newblock ``Estimating the uncertainty in emotion attributes using deep
  evidential regression,''
\newblock in {\em Proceedings of the 61st Annual Meeting of the Association for
  Computational Linguistics (Volume 1: Long Papers)}, 2023, pp. 15681--15695.

\bibitem{gui2024active}
Shurui Gui, Xiner Li, and Shuiwang Ji,
\newblock ``Active test-time adaptation: Theoretical analyses and an
  algorithm,''
\newblock in {\em International Conference on Learning Representations (ICLR)},
  2024.

\bibitem{zhang2015cooperative}
Zixing Zhang, Eduardo Coutinho, Jun Deng, and Bj{\"o}rn Schuller,
\newblock ``Cooperative learning and its application to emotion recognition
  from speech,''
\newblock {\em IEEE/ACM Transactions on Audio, Speech, and Language
  Processing}, vol. 23, no. 1, pp. 115--126, 2015.

\bibitem{abdelwahab2017incremental}
Mohammed Abdelwahab and Carlos Busso,
\newblock ``Incremental adaptation using active learning for acoustic emotion
  recognition,''
\newblock in {\em Proc. IEEE International Conference on Acoustics, Speech and
  Signal Processing (ICASSP)}, 2017, pp. 5160--5164.

\bibitem{sperber2013respeaking}
Matthias Sperber, Graham Neubig, Christian F{\"u}gen, Satoshi Nakamura, and
  Alex Waibel,
\newblock ``Efficient speech transcription through respeaking,''
\newblock in {\em Proc. Interspeech 2013}, 2013, pp. 1087--1091.

\bibitem{sperber2014segmentation}
Matthias Sperber, Mirjam Simantzik, Graham Neubig, Satoshi Nakamura, and Alex
  Waibel,
\newblock ``Segmentation for efficient supervised language annotation with an
  explicit cost-utility tradeoff,''
\newblock {\em Transactions of the Association for Computational Linguistics},
  vol. 2, pp. 169--180, 2014.

\bibitem{shamsi2022diarization}
Meysam Shamsi, Anthony Larcher, Lo{\"i}c Barrault, Sylvain Meignier, Yevhenii
  Prokopalo, Marie Tahon, Ambuj Mehrish, Simon Petitrenaud, Olivier Galibert,
  Samuel Gaist, Andr{\'e} Anjos, S{\'e}bastien Marcel, and Marta~R.
  Costa-juss{\`a},
\newblock ``Towards lifelong human assisted speaker diarization,''
\newblock {\em Computer Speech \& Language}, vol. 77, pp. 101437, 2023.

\bibitem{kalman1960}
R.~E. Kalman,
\newblock ``A new approach to linear filtering and prediction problems,''
\newblock {\em Journal of Basic Engineering}, vol. 82, no. 1, pp. 35--45, 1960.

\bibitem{durbin2012}
James Durbin and Siem~Jan Koopman,
\newblock {\em Time Series Analysis by State Space Methods},
\newblock Oxford University Press, Oxford, 2nd edition, 2012.

\bibitem{rauch1965}
H.~E. Rauch, F.~Tung, and C.~T. Striebel,
\newblock ``Maximum likelihood estimates of linear dynamic systems,''
\newblock {\em AIAA Journal}, vol. 3, no. 8, pp. 1445--1450, 1965.

\bibitem{mariooryad2015delay}
Soroosh Mariooryad and Carlos Busso,
\newblock ``Correcting time-continuous emotional labels by modeling the
  reaction lag of evaluators,''
\newblock {\em IEEE Transactions on Affective Computing}, vol. 6, no. 2, pp.
  97--108, 2015.

\bibitem{chen2018neuralode}
Ricky T.~Q. Chen, Yulia Rubanova, Jesse Bettencourt, and David~K. Duvenaud,
\newblock ``Neural ordinary differential equations,''
\newblock in {\em Advances in Neural Information Processing Systems}, 2018,
  vol.~31.

\bibitem{chen2022wavlm}
Sanyuan Chen, Chengyi Wang, Zhengyang Chen, Yu~Wu, Shujie Liu, Zhuo Chen, Jinyu
  Li, Naoyuki Kanda, Takuya Yoshioka, Xiong Xiao, Jian Wu, Long Zhou, Shuo Ren,
  Yanmin Qian, Yao Qian, Jian Wu, Michael Zeng, Xiangzhan Yu, and Furu Wei,
\newblock ``{WavLM}: Large-scale self-supervised pre-training for full stack
  speech processing,''
\newblock {\em IEEE Journal of Selected Topics in Signal Processing}, vol. 16,
  no. 6, pp. 1505--1518, 2022.

\bibitem{burmania2017raters}
Alec Burmania and Carlos Busso,
\newblock ``A stepwise analysis of aggregated crowdsourced labels describing
  multimodal emotional behaviors,''
\newblock in {\em Proc. Interspeech 2017}, 2017, pp. 152--156.

\bibitem{cochran1977}
William~G. Cochran,
\newblock {\em Sampling Techniques},
\newblock John Wiley \& Sons, New York, 3rd edition, 1977.

\bibitem{lin1989ccc}
Lawrence I-Kuei Lin,
\newblock ``A concordance correlation coefficient to evaluate
  reproducibility,''
\newblock {\em Biometrics}, vol. 45, no. 1, pp. 255--268, 1989.

\bibitem{field2007bootstrap}
C.~A. Field and A.~H. Welsh,
\newblock ``Bootstrapping clustered data,''
\newblock {\em Journal of the Royal Statistical Society: Series B (Statistical
  Methodology)}, vol. 69, no. 3, pp. 369--390, 2007.

\end{thebibliography}

\end{document}